\documentclass[a4paper]{article}
\usepackage{graphicx}

\begin{document}

\begin{flushright} {\large SAGA-HE-147} \end{flushright}

\vspace{1cm}

\begin{center}
{\huge Dispersion relations in quantum hadrodynamics with a finite cutoff}
\end{center}

\centerline{Katsuaki Sakamoto$^a$, Hiroaki Kouno, Akira Hasegawa
and Masahiro Nakano$^*$}

\centerline{Department of Physics, Saga University Saga 840, Japan}

\centerline{*University of Occupational and Environmental Health, Kitakyushu 807, Japan}

\vspace{5mm}

\centerline{$^a$ E-mail: 98td22@edu.cc.saga-u.ac.jp}

\vspace{5mm}

\centerline{\bf Abstract}

The dispersion relations in the real and imaginary parts of the meson self-energies are studied to check the consistency of the "renormalization" in cutoff field theory. 
It is shown that the dispersion relations are preserved by the "renormalization" 
even if the finite cutoff and regulator are introduced in the calculation by hand.

\section{Introduction}

In the past two decades, nuclei and nuclear matter have been studied in the framework of quantum hadrodynamics (QHD).\cite{rf:Walecka,rf:Serot} 
The meson mean-field theory of nuclear matter\cite{rf:Walecka} has produced successful results to account for the saturation properties at normal nuclear density. 
Following these successes, many studies and modifications have been performed on relativistic nuclear models. 
One of these modifications is the inclusion of vacuum fluctuation effects, which cause divergences in physical quantities when they are naively calculated. 
Chin\cite{rf:Chin} estimated vacuum fluctuation effects in the Hartree approximation by using a renormalization procedure, and found that vacuum fluctuation effects make the incompressibility of nuclear matter smaller and closer to the empirical value than in the original Walecka model. 

Although the relation between QHD and the underlying fundamental theory, i.e., QCD, is an open question, it is natural that QHD is not valid at very high-energy. 
In this point of view, a cutoff or a form factor should be introduced into the theory of QHD. 
One may introduce the cutoff\cite{rf:Kohmura} or the form factor\cite{rf:Furn} to avoid the instability of the meson propagators in the random phase approximation (RPA)\cite{rf:Kurasawa,rf:Lim}. 
Cohen\cite{rf:Cohen} introduced a four-dimensional cutoff into the relativistic Hartree calculation and found that the vacuum energy contribution may be somewhat different from that in the ordinary renormalization procedures, if the cutoff is not so large. 

One troublesome problem in use of the cutoff is that physical results depend on a value of the cutoff and a form of the regulator which are introduced into the theory by hand. 
It is difficult to determine a suitable value of cutoff and a appropriate shape of the regulator phenomenologically. 
For the high-energy physics near the limitation of the theory of the particle physics, Lepage\cite{rf:Lepage} proposed to use the extended effective action (Lagrangian) to eliminate such a dependence on cutoff which is by hand introduced into the theory, rather than to search the phenomenologically favorable cutoff and regulator (or form factor). 
This idea is based on the renormalization in the non-perturbative renormalization group (NPRG) method. 
\cite{rf:Wilson}-\cite{rf:Kouno4}. 
In the series of papers\cite{rf:Kouno1,rf:Kouno2,rf:Kouno3}, we have studied nuclear matter properties and the vertex corrections in the framework of the cutoff field theory, eliminating the cutoff dependence by using the Lepage's method. 

Although the utility of our method of renormalization has been checked phenomenologically \cite{rf:Kouno2}, 
one may wonder the finite cutoff can be introduced consistently and does not break the consistent condition. 
In this paper, we study the dispersion relation between real and imaginary parts of the meson self-energy in the framework of the finite cutoff field theory to check the consistency of our method and show how the consistency is preserved in our formalism. 

This paper is organized as follows. 
In {\S}2, we review and reformulate our method to remove the cutoff dependences of the physical results. 
In {\S}3, we study the dispersion relation between real and imaginary parts of the meson self-energy in the framework of the finite cutoff field theory to check the consistency of our method. 
In {\S}4, the influence of the finite cutoff effects to the vacuum polarization is discussed. 
Section 5 is devoted to a summary. 

\section{Vacuum polarization effects in cutoff field theory}

In this section we review the "renormalizations" based on cutoff field theory, discussing the vacuum polarization effects. 
After that, we reformulate the method in the form of differential equations.

At first, we start from " renormalizable " Lagrangian which defined at cutoff $\Lambda_0$. 
The cutoff $\Lambda_0$ denotes a upper limit energy scale below which the theory can be used. 
We will show that "nonrenormalizable" terms will appear in the low-energy effective Lagrangian, even if we start with a "renormalizable" Lagrangian. 
We demonstrate the Lagrangian does not have to be "renormalizable" in the sense of the effective action and $should$ have the "non-renormalizable" terms at low-energy. 

The "renormalizable" Lagrangian of the $\sigma$-$\omega$ model is
\begin{eqnarray}
L_{\Lambda_0} &=& \bar{\psi}(i \gamma_{\mu} \partial^{\mu}-M+g_s \phi-g_v
\gamma_{\mu}V^{\mu}) \psi
+\frac{1}{2}\partial_{\mu} \phi \partial^{\mu} \phi-\frac{1}{2}m_s^2\phi^2
+\frac{1}{2}F_1\partial_{\mu} \phi \partial^{\mu} \phi-\frac{1}{2}F_0\phi^2
\nonumber \\
 & & -U(\phi)-\frac{1}{4}F_{\mu \nu}F^{\mu \nu}+\frac{1}{2}m_v^2V_{\mu}V^{\mu},
\nonumber \\
U(\phi) &=& \sum_{n=0}^4 C_n(g_s \phi)^n,
\label{eq:l}
\end{eqnarray}
where $\psi$, $\phi$, $V_{\mu}$, $M$, $m_s$, $m_v$, $g_s$ and $g_v$
are the nucleon field, $\sigma$-meson field, $\omega$-meson field,
nucleon mass, $\sigma$-meson mass, $\omega$-meson mass,
$\sigma$-nucleon coupling and $\omega$-nucleon coupling, respectively.
The vector field strength is $F_{\mu \nu}=\partial_{\mu}V_{\nu}-\partial_{\nu}V_{\mu}$, 
and $C_n$ are "bare" $\sigma$-meson self-interaction couplings which are 
adjusted to reproduce renormalization conditions. 
For simplicity, in Eq. (\ref{eq:l}) we omitted the other counter-terms which are not necessary for the relativistic Hartree approximation(RHA) and the random phase approximation (RPA) for $\sigma$-meson self-energy. 
The Lagrangian (\ref{eq:l}) is valid only in the low-energy region below $\Lambda_0$.

For nucleon propagations, we use the solution in RHA. 
In this approximation the meson fields are replaced
by their expectation values, 
\begin{eqnarray}
\phi &\longrightarrow& \langle \phi \rangle \equiv \phi_0 \nonumber \\
V_\mu &\longrightarrow& \langle V_\mu \rangle \equiv \delta_{\mu 0}V_0.
\label{eq:2}
\end{eqnarray}
At zero-density, the nucleon propagator $G(p)$ is just the same as the Feynman propagator for free nucleon, i.e., 
\begin{eqnarray}
G(p)={{\gamma_\mu p^\mu +M}\over{p^2-M^2+i\epsilon }} 
\label{eq:3}
\end{eqnarray}
 
Let us estimate one-loop vacuum polarization in the RPA, using this nucleon propagator.  
For simplicity of the illustration, in this section, we consider vacuum polarization at zero baryon density. 
The extension to the case of the finite baryon density is straightforward 
as in the ordinary renormalization procedure. \cite{rf:Chin} 
We will discuss the RPA calculations at finite density in the next section.

\begin{figure}[p]
\begin{center}
    \includegraphics*[height=5cm]{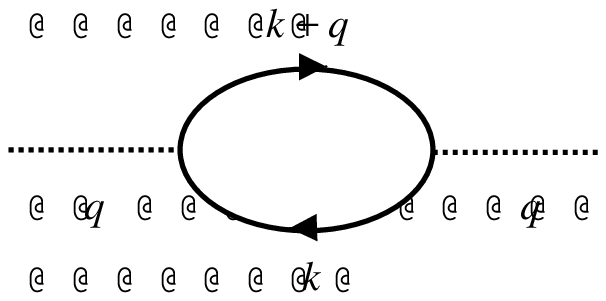}
\caption{
Feynman diagram for RPA meson self-energy. The solid line
denotes the nucleon and the dotted lines denote the meson.
}
\label{f}
\end{center}
\end{figure}

If we use the sharp regulator with the cutoff $\Lambda_0$, one-loop RPA contribution of the $\sigma$-meson self-energy (Fig. \ref{f}) is given by
\begin{eqnarray}
\Pi(q^2,\Lambda_0)
=\frac{g_s^2}{2 \pi^2} \int_0^1dx \int_0^{\Lambda_0}
dk^\prime \frac{k^{\prime 3}(-k^{\prime 2}+A^2)}{(k^{\prime 2}+A^2)^2},
\label{eq:pi}
\end{eqnarray}
where $A^2=M^2-q^2x(1-x)$, $k^\prime=k-q(1-x)$.

\vspace{1cm}
\begin{center}
  \begin{tabular}{c} \hline
     Fig.\ref{f} \\ \hline
  \end{tabular}
\end{center}
\vspace{1cm}

In Eq. (\ref{eq:pi}), the cutoff $\Lambda_0$ appears.
Suppose that we do not know a correct value of $\Lambda_0$ which means 
a limit of an energy scale of the theory and we use another cutoff $\Lambda$
which is a smaller than $\Lambda_0$ in the calculation. 
(Of course, $\Lambda$ should be larger than the energy scale of the physics
in which we are interested.)

As in the renormalization group equations, we require that physical quantities do not depend on $\Lambda$, although the physical quantities depend on $\Lambda_0$. 
To achieve this, we estimate the contribution which is needlessly discarded
by using the cutoff $\Lambda$ instead of $\Lambda_0$.
In the case of $\sigma$-meson self-energy (\ref{eq:pi}), this is given by
\begin{eqnarray}
\Delta \Pi(q^2,\Lambda_0,\Lambda)
 &=& \Pi(q^2,\Lambda_0)-\Pi(q^2,\Lambda)
\nonumber \\
 &=& \frac{g_s^2}{2 \pi^2}
\int_0^1dx \int_\Lambda^{\Lambda_0}dk^\prime 
\frac{k^{\prime 3}(-k^{\prime 2}+A^2)}{(k^{\prime 2}+A^2)^2}
\nonumber \\
 &=& \frac{g_s^2}{4 \pi^2}\Lambda^2
\int_0^1dx \int_1^ady
\frac{y[-y+z-ux(1-x)]}{[y+z -ux(1-x)]^2},
\label{eq:pi2}
\end{eqnarray}
where $y=k^{\prime 2}/\Lambda^2$, $z=M^2/\Lambda^2$, $u=q^2/\Lambda^2$
and $a=\Lambda_0^2/\Lambda^2$.
If external momentum $q$ is smaller than $\Lambda$, we could expand the
Eq. (\ref{eq:pi2}) around $u=0$ and we get
\begin{eqnarray}
\Delta \Pi(q^2,\Lambda_0,\Lambda)
 &=& \sum^\infty_{n=0} \frac{1}{n!} \frac{d^n}{du^n}\Delta \Pi|_{u=0}u^n
\nonumber \\
 &\equiv & \sum^\infty_{n=0}D_n^*\Lambda^2(q^2/ \Lambda^2)^n.
\nonumber \\
 &\equiv & \sum^\infty_{n=0}D_n(q^2)^n.
\label{eq:pi3}
\end{eqnarray}
Each term of expansion (\ref{eq:pi3}) corresponds to a quantity which
is proportional to $(\partial^2)^n \phi^2$ in Lagrangian.
Therefore, if we add terms, which is equivalent to (\ref{eq:pi3}), to Lagrangian at the beginning of our calculation and determine them phenomenologically, we do not need to know about $\Lambda_0$. 
In that way, we get proper physical results by using the cutoff $\Lambda$ instead of $\Lambda_0$.  

In other word, this means that the $\Lambda$-dependence of the physical results is removed by the phenomenological determinations of the new terms, since $\Lambda$ is introduced by hand and has no physical meaning.  
To say exactly, we must determine all coefficients of expansion (\ref{eq:pi3}) phenomenologically, to eliminate all $\Lambda$-dependence. 
It is not possible actually. 
However, if $\Lambda$ is the same order as $\Lambda_0$ and if $q^2$ is smaller than $\Lambda^2$, it is easily seen that $D_n^*$ is of order 1 and the n-th term in the expansion (\ref{eq:pi3}) is order of $(1/\Lambda^2)^{n-1}$. 
Therefore, if we consider a limit of $\Lambda \rightarrow \infty$, the terms $D_n(q^2)$ $(n \ge 2)$ in (\ref{eq:pi3}) amounts to zero, and we need to determine only the coefficients $D_n(n \le 1)$ phenomenologically. 
In this limit, it is clear that this procedure is essentially the same as the usual renormalization procedure in which the coefficient $F_0$ is chosen to renormalize the $\sigma$-meson mass and the coefficient $F_1$ is chosen to renormalize the $\sigma$-meson wave function. 
We must determine $F_0+D_0$ and $F_1+D_1$ instead of $F_0$ and $F_1$, if we use $\Lambda$ instead of $\Lambda_0$. 
However, the number of the phenomenological inputs is unchanged, since we only need to know the sum $F_i+D_i$ and do not need to know $F_i$ and $D_i$, respectively. 
In this meaning, the effects of quantum fluctuations are "renormalizable" if we re-define the counter-terms as $F_0^*=F_0+D_0$ and $F_1^*=F_1+D_1$. 

However, since the cutoff $\Lambda_0$ may be several GeV in QHD, we can not take a limit $\Lambda \rightarrow \infty$. 
In this case, the errors of order $(1/\Lambda)^2$ may not be negligible. 

According to Lepage's proposal, we determine not only the coefficient $C_n+D_n~(n \le 1)$ but also the higher coefficients  $D_n~( n>1)$ phenomenologically to remove the errors which occurs from the finiteness of $\Lambda$. 
As an example, we determine the coefficients $D_2$ phenomenologically in addition to $F_0^*$ and $F_1^*$, 
if we want results with $O(1/\Lambda^4)$ errors. 
This is equivalent to add the terms, which are proportional to $(\partial^2)^2 \phi^2$, to the Lagrangian. 
Moreover, if we want the errors which are $O((1/\Lambda^2)^{N})$, 
we should use the effective Lagrangian 
\begin{eqnarray}
L_\Lambda=L_{\Lambda_0}+\sum^N_{n=0}{1\over{2}}D_n(\partial^2)^n \phi^2. 
\label{eq:el}
\end{eqnarray}
with the cutoff $\Lambda$. 
Similarly, adding higher terms to the Lagrangian and determining them phenomenologically, the $\Lambda$-dependence of the physical results  are removed order by order. 

It should also be emphasized that the low-energy effective Lagrangian such as (\ref{eq:el}) has "nonrenormalizable" terms, although the original Lagrangian (\ref{eq:l}) is "renormalizable". 
Of course, they cause no problem, because of the coefficients of "nonrenormalizable" terms contain a suppression factor $(1/\Lambda^2 )^{n-1}$. 
If we use the "nonrenormalizable" Lagrangian (\ref{eq:el}) with $\Lambda$ instead of the "renormalizable " Lagrangian (\ref{eq:l}) with the cutoff $\Lambda_0$, we get the proper results. 

We also remark that the $\Lambda_0$-dependence of the effective Lagrangian is almostly included in the terms which are determined phenomenologically. 
Therefore, we do not need detailed knowledge of cutoff $\Lambda_0$. 

In the discussions above, we assume the Lagrangian is "renormalizable" at $\Lambda_0$. 
However, the discussions above indicate that the Lagrangian may have following "nonrenormalizable terms" even at $\Lambda_0$. 
\begin{eqnarray}
L_{\Lambda_0}^\prime =L_{\Lambda_0}+\sum^{\infty}_{n=2}{1\over{2}}F_n(\partial^2)^n\phi^2, 
\label{eq:Ad1117b}
\end{eqnarray}
where $F_n$ is order of $(1/\Lambda_0^2)^{n-1}$. 
In this case, we need to determine only $F_n+D_n~(n\leq N)$ and put $F_n+D_n=0~(n>N)$ if we want results with errors of order $(1/\Lambda^2)^{N}$. 
The number of the phenomenological inputs is unchanged as before. 
Therefore, in general, we consider the extended Lagrangian (\ref{eq:Ad1117b}) at the initial. 

Next we reformulate the results above by using a differential equation
in the finite cutoff theory. \cite{rf:Aoki4,rf:Weinberg} 
In this formulation, it is more clear that the physical results hardly depend on the regulator which was introduced by hand. 

Here we consider $\sigma$-meson self-energy (\ref{eq:pi}) again, using the cutoff $\Lambda$ instead of the true one $\Lambda_0$. 
The $\Lambda$-dependence of $\Pi (q^2, \Lambda^2 )$ is determined by the high-energy behaviour of the integrand. 
Therefore, we can easily see that $\Pi$ has the $\Lambda$-dependence of order 
$\Lambda^2$ by power counting. 
The ambiguity of $O(\Lambda^2)$ is too large, since $\Lambda$ is larger than the physical energy scale of the problem. 
Therefore, instead of $\Pi$ itself, we consider a second derivative  
\begin{eqnarray}
\Pi^{(2)}(q^2,\Lambda)=\frac{d^2}{(dq^2)^2}\Pi(q^2,\Lambda),
\label{eq:dp}
\end{eqnarray}
of it. 
We can know that Eq. (\ref{eq:dp}) is $O(1/\Lambda^2)$ by power counting. 
If $\Lambda$ is sufficiently large, this means that $\Pi^{(2)}$ hardly depends on $\Lambda$. 
Especially, in the the usual renormalization procedure ($\Lambda \rightarrow \infty$), $\Pi^{(2)}$ is independent of $\Lambda$. 

To get the self-energy $\Pi$ itself, we regard Eq. (\ref{eq:dp}) as a differential equation and integrate it. \cite{rf:Aoki4,rf:Weinberg} 
We get  
\begin{eqnarray}
\Pi_2(q^2,\Lambda)=F_0^\prime +F_1^\prime q^2+ \int_c^sdt \int_c^tdv \Pi^{(2)}(v,\Lambda),
\label{eq:ip}
\end{eqnarray}
where $C_0$ and $C_1$ are integration constants, and $c$ denotes a "renormalization point". 
If $C_0$ and $C_1$ are determined phenomenologically, the $\Lambda$-dependence of $\Pi_2$ is order of $1/\Lambda^2$, and if $\Lambda$ is sufficiently large, we can ignore the dependence. 
But we may not be able to ignore $\Lambda$-dependence of $\Pi_2$, if $\Lambda$ is not so large.
In this case, we need to repeat the differentiations until the $\Lambda$-dependence becomes sufficiently small. 
If we differentiate $\Pi$ $(N+1)$-times with respect to $q^2$, we get 
\begin{eqnarray}
\Pi^{(N+1)}(q^2,\Lambda)=\frac{d^{N+1}}{(dq^2)^{N+1}}\Pi(q^2,\Lambda)
\qquad (N+1>2).
\label{eq:dp2}
\end{eqnarray}
The $\Lambda$-dependence of Eq. (\ref{eq:dp2}) is order of $(1/\Lambda^2)^N$.
We integrate Eq. (\ref{eq:dp2}) as in the case of Eq. (\ref{eq:dp}). 
We get 
\begin{eqnarray}
\Pi_{N+1}(q^2,\Lambda) &=& \sum_{n=0}^N F^\prime_n(q^2)^n \nonumber \\
 & & +\int_c^sds_1 \int_c^{s_1}ds_2 \cdot \cdot \cdot \cdot
\int_c^{s_N}ds_{N+1} \Pi^{(N+1)}(s_{N+1},\Lambda).
\label{eq:ip2}
\end{eqnarray}
If $F_0^\prime \sim F_N^\prime $ are determined phenomenologically, the $\Lambda$-dependence of $\Pi_{N+1}$ is order of $(1/\Lambda^2)^N$. 

Therefore, if $\Lambda$-dependence of physical quantities are not small enough to be ignored, we should differentiate the physical quantities by some external parameter until the $\Lambda$-dependence becomes small enough.  
It is clear that this formulation is equivalent to the one proposed by Lepage. 
Furthermore, in this reformulation, we easily see that our formulation does not depend on the details of the regulator, since we have not assume its explict form. 
This means that the phenomenological determinations of the coefficients of the effective Lagrangian remove the dependence on the form of the regulator which is introduced by hand, as well as the $\Lambda$-dependence.

\section{Dispersion relations in $\sigma$-$\omega$ self-energies}

In this section, we ascertain dispersion relations of RPA $\sigma$-$\omega$
self-energies in finite density system.

We study the RPA meson self-energies in finite density system. 
The $\sigma$-meson self-energy, the $\omega$-meson self-energy, the $\sigma$-$\omega$ 
mixed polarization part self-energy are given by
\begin{eqnarray}
\Pi_{s}(q) &=& -ig^{2}_{s}\int\frac{d^{4}k}{(2\pi)^4}Tr[G(k)G(k+q)], 
\nonumber \\
\Pi_{\mu\nu}(q) &=& -ig^{2}_{v}\int\frac{d^{4}k}{(2\pi)^4}
Tr[\gamma_{\mu}G(k)\gamma_{\nu}G(k+q)], \nonumber \\
\Pi_{\mu}^{M}(q) &=& ig_{s}g_{v}\int\frac{d^{4}k}{(2\pi)^4}
Tr[\gamma_{\mu}G(k)G(k+q)],
\label{eq:fpi}
\end{eqnarray}
where $G$ is nucleon propagator in RHA and is given by
\begin{eqnarray}
G(k) &=& (\gamma^{\mu}k^{\ast}_{\mu}+M^{\ast})
\Bigl[\frac{1}{k^{\ast2}-M^{\ast2}+i\epsilon}
+\frac{i\pi}{E^{\ast}(k)} \delta(k_{0}^{\ast}
-E^{\ast}(k))\theta(k_{F}-|\mbox{\boldmath $k$}|) \Bigr] \nonumber \\
&\equiv& G_{F}(k)+G_{D}(k),
\label{eq:g}
\end{eqnarray}
where $k_F$ is the Fermi momentum. 
The $M^{\ast}$ is the effective nucleon mass which is related with the vacuum expectation value $\phi_0$ of $\sigma$-meson field as 
\begin{eqnarray}
M^*=M-g_s \phi_0.
\label{eq:em}
\end{eqnarray}
The effective three dimensional momentum and the effective energy of the nucleon are given by 
\begin{eqnarray}
k^{\ast\mu} &=& (k^0-g_vV^0,\mbox{\boldmath $k$}), \nonumber \\
E^{\ast}(k) &=& \sqrt{\mbox{\boldmath $k$}^2+M^{\ast 2}}.
\label{eq:mo}
\end{eqnarray}
In Eq. (\ref{eq:g}), $G_F$ describes the propagation of nucleon and antinucleon. 
The $G_D$ describes the propagation of holes in the Fermi sea and it also include the effects of the Pauli exclusion principle for the particle and antiparticle excitations below the Fermi surface. 

Here we substitute Eq. (\ref{eq:g}) into Eq. (\ref{eq:fpi}),
we can divide meson self-energy into a vacuum fluctuation part and a density dependent part. 
\begin{eqnarray}
\Pi(q)=\Pi^V(q)+\Pi^D(q).
\label{eq:sp}
\end{eqnarray}
The density dependent part and vacuum fluctuation part are given by 
\begin{eqnarray}
\Pi^D_{s}(q) &=& -ig^{2}_{s}\int\frac{d^{4}k}{(2\pi)^4}
Tr[G_D(k)G_D(k+q)+G_D(k)G_F(k+q) \nonumber \\
& &+G_F(k)G_D(k+q)], \nonumber \\
\Pi^D_{\mu\nu}(q) &=& -ig^{2}_{v}\int\frac{d^{4}k}{(2\pi)^4}
Tr[\gamma_{\mu}G_D(k)\gamma_{\nu}G_D(k+q) \nonumber \\
& &+\gamma_{\mu}G_D(k)\gamma_{\nu}G_F(k+q)+\gamma_{\mu}G_F(k)
\gamma_{\nu}G_D(k+q)], \nonumber \\
\Pi_{\mu}^{M}(q) &=& ig_{s}g_{v}\int\frac{d^{4}k}{(2\pi)^4}
Tr[\gamma_{\mu}G(k)G(k+q)], 
\label{eq:PiD}
\end{eqnarray}
and
\begin{eqnarray}
\Pi^V_{s}(q) &=& -ig^{2}_{s}\int\frac{d^{4}k}{(2\pi)^4}Tr[G_F(k)G_F(k+q)], 
\nonumber \\
\Pi^V_{\mu\nu}(q) &=&
-ig^{2}_{v}\int\frac{d^{4}k}{(2\pi)^4}
Tr[\gamma_{\mu}G_F(k)\gamma_{\nu}G_F(k+q)]\nonumber \\
 &=& \Bigl( \frac{q_\mu q_\nu}{q^2}-g_{\mu \nu} \Bigr) \Pi^V_v(q). 
\label{eq:PiV}
\end{eqnarray}
We easily see that the vacuum fluctuation part of the $\sigma$-meson self-energy is equal to the one at zero-density (Eq. (\ref{eq:pi})) if we use effective nucleon mass $M^*$ instead of $M$. 
This is true for the vacuum fluctuation part of the $\omega$-meson self-energy. 
Therefore, as well as in the case of zero-baryon density, the vacuum fluctuation part $\Pi^V$ is divergent when it is calculated naively. 
So we introduce the cutoff $\Lambda$ to regularize the divergences and perform the "renormalization" procedure described in the previous section, i.e., in vacuum fluctuation part, we examine a dispersion relations using the derivative of $\Pi^V$ rather than $\Pi^V$ itself. 
In this section, we adopt a limit of $\Lambda \rightarrow \infty$. 
The effects of the finiteness of the cutoff to the dispersion relation will be discussed in the next section. 

First we consider dispersion relations in the density part $\Pi^D$. 
On an energy complex plane, it is given by 
\begin{eqnarray}
Re\Pi^{D}(q)=\frac{2{\rm \bf P}}{\pi}\int_0^\infty
\frac{Im\Pi^{D}(\omega,\mbox{\boldmath $q$})\cdot\omega}
{\omega^2-q_0^2}d\omega,
\label{eq:dd}
\end{eqnarray}
where "{\bf P}" denotes the principal value. 
The derivation of Eq. (\ref{eq:dd}) is presented in Appendix. 

In Fig. \ref{dd}, we show dispersion relations of density part Eq. (\ref{eq:dd}). 
In this figure, the results calculated directly by using Eq. (\ref{eq:PiD}) are shown by solid, dotted, dashed, dashed-dotted lines, while the corresponding results calculated by using the dispersion relations Eq. (\ref{eq:dd}) are shown by filled circles. 
We have put the absolute value of the three dimensional momentum $q=100MeV$ in the calculation. 
We see that each line agrees with filled circles. 
The dispersion relations are preserved in the density part. 

\vspace{1cm}
\begin{center}
  \begin{tabular}{c} \hline
     Fig.\ref{dd} \\ \hline
  \end{tabular}
\end{center}
\vspace{1cm}

We assume the same relation in the vacuum fluctuation part $\Pi^V$. 
However $\Pi^V$ has strong cutoff dependence and diverges in the limit $\Lambda \rightarrow \infty$. 
Samely, the large $\omega$ contributions do not vanish in the $\omega$-integral in the dispersion relation and the $\omega$-integral diverges. 
So we can not apply the dispersion relations to $\Pi^V$ just as it is. 
Therefore we consider the dispersion relation in the second derivative of $\Pi^{V}$ rather than $\Pi^V$ itself. 
\begin{eqnarray}
\frac{d^2}{(dq_0^2)^2}[Re\Pi^{V}(q)] = \frac{d^2}{(dq_0^2)^2}
\Bigl[\frac{2{\rm \bf P}}{\pi}\int_0^\infty
\frac{Im\Pi^{V}(\omega,\mbox{\boldmath $q$})\cdot\omega}
{\omega^2-q_0^2}d\omega \Bigr].
\label{eq:dv}
\end{eqnarray}

\begin{figure}
\begin{center}
    \includegraphics*[height=8cm]{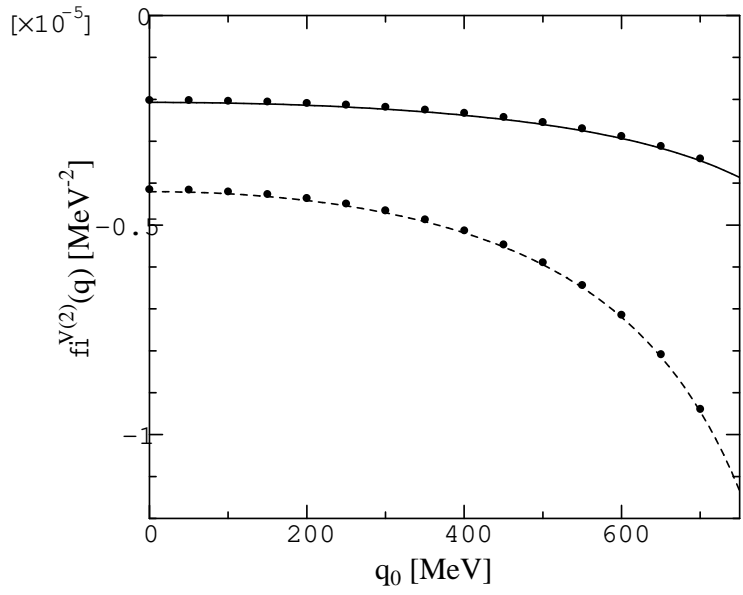}
\caption{
Dispersion relation for the second derivatives of the vacuum fluctuation part. 
The solid, dashed lines are the second derivatives of $Re\Pi^V_s$ 
and $Re\Pi^V_v$, respectively. 
The corresponding results calculated by Eq. (\ref{eq:dv}) are shown by the filled circles. 
}
\label{dv}
\end{center}
\end{figure}

The numerical calculations for the dispersion relation Eq. (\ref{eq:dv}) in the vacuum fluctuation part are shown in Fig. \ref{dv}. 
We see that Eq. (\ref{eq:dv}) is preserved as well as Eq. (\ref{eq:dd}). 
The dispersion relation is preserved in the conversing second derivative of the diverging quantity. 

\vspace{1cm}
\begin{center}
  \begin{tabular}{c} \hline
     Fig.\ref{dv} \\ \hline
  \end{tabular}
\end{center}
\vspace{1cm}

Here we add a comment which concerns the result of Fig. \ref{dv}. 
When we calculated $\Pi^V$ using Eq. (\ref{eq:PiV}) , we have used the Wick rotation. 
Similar as the dispersion relation, one condition for the use of the Wick rotation is that the integrand of the energy integral in $\Pi^V$ should not contribute in the limit where a complex variable go to $\infty$. 
However, this condition is not satisfied and $\Pi^V$ diverges. 
This means that, to say exactly, we cannot directly apply the Wick rotation to $\Pi^V$. 
However, we know that we can get the justifiable physical results by using the renormalization procedure after the Wick rotation. 
In other words, the ambiguities which have appeared in the use of the Wick rotation is canceled by the renormalization procedure. 
However, the results above indicate that, alternatively, we can consider the conversing derivative of the diverging quantity before we use the Wick rotation. 
In fact, this is more clear interpretation for the Wick rotation. 
Furthermore, as is shown in Appendix, the Wick rotation in $k_0$-plane can be consistent with the dispersion relation in $\omega$-plane. 
We can interpret the Wick rotation and the dispersion relation on an equal footing, by considering the conversing derivatives of the diversing quantities.

\section{The effects of the finite cutoff to the vacuum polarization}

Next, we study the cutoff dependence in $\sigma$ and $\omega$ self-energies numerically, and we examine the effects of the finiteness of the cutoff in the dispersion relations Sec.3. 
After that, we discusses the renormalization in QHD which is defined by
the cutoff which is a order of several GeV. 

We examine the finite cutoff dependence in the derivatives of $\Pi^V$ numerically. 
Consider second and third derivatives of the vacuum parts of the meson self-energies with respect to $q^2$. 
We call them $\Pi^{V(2)}$ and $\Pi^{V(3)}$ respectively. 
The $\Lambda$-dependence of $\Pi^{V(2)}$ are order of ($1/\Lambda^2$), while the $\Lambda$-dependence of $\Pi^{V(3)}$ are order of
($1/\Lambda^4$). 
Naturally, in the ordinary renormalization procedure, the limit, $\Lambda \rightarrow \infty$, is taken, and the cutoff dependences of $\Pi^{V(2)}$ and $\Pi^{V(3)}$ vanish.  
In the case of the finite cutoff, we should check an utility of a method of renormalization in Sec.2 by numerical calculations. 

\begin{figure}
\begin{center}
    \includegraphics*[height=16cm]{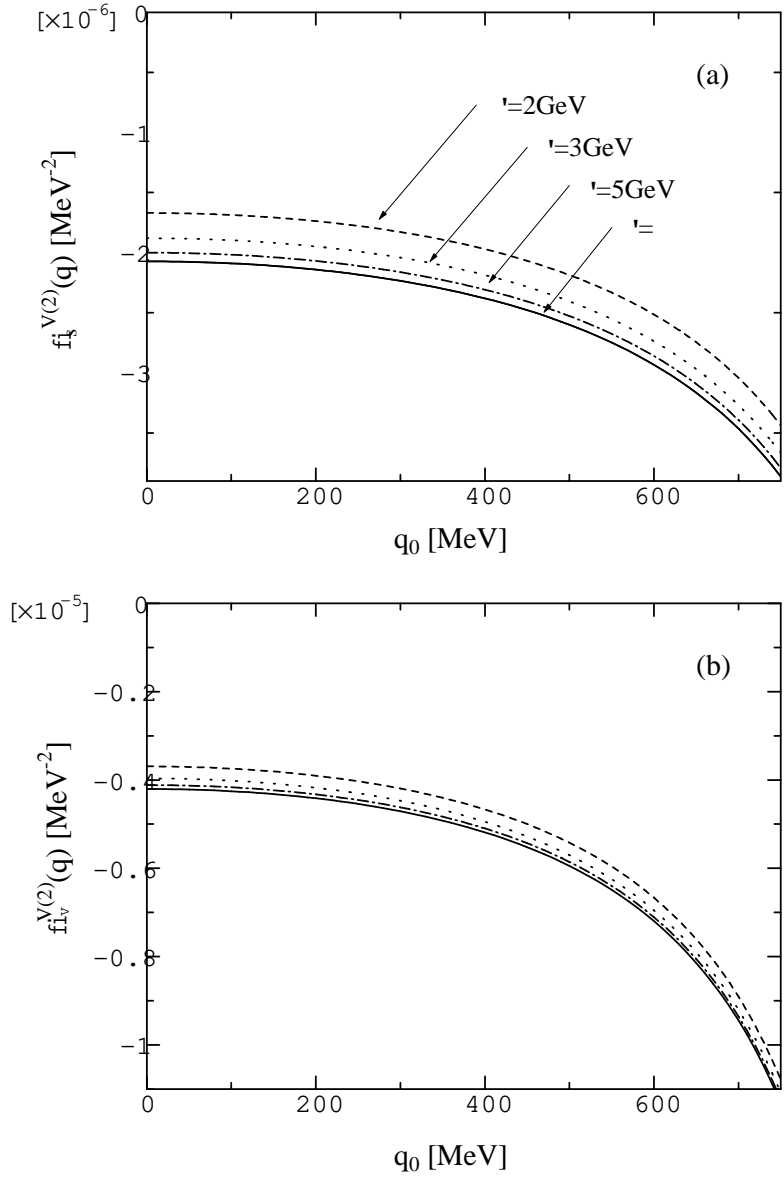}
\caption{
The $\Lambda$-dependence of the second derivative of $\Pi^V$. 
The dashed, dotted, dashed-dotted and solid line are the results with $\Lambda$ = 2 GeV, $\Lambda$ = 3 GeV, $\Lambda$ = 5 GeV and $\Lambda$ = $\infty$, respectively. 
(a) The $\sigma$-meson self-energy, 
(b) The $\omega$-meson self-energy. 
}
\label{2}
\end{center}
\end{figure}

\begin{figure}
\begin{center}
    \includegraphics*[height=16cm]{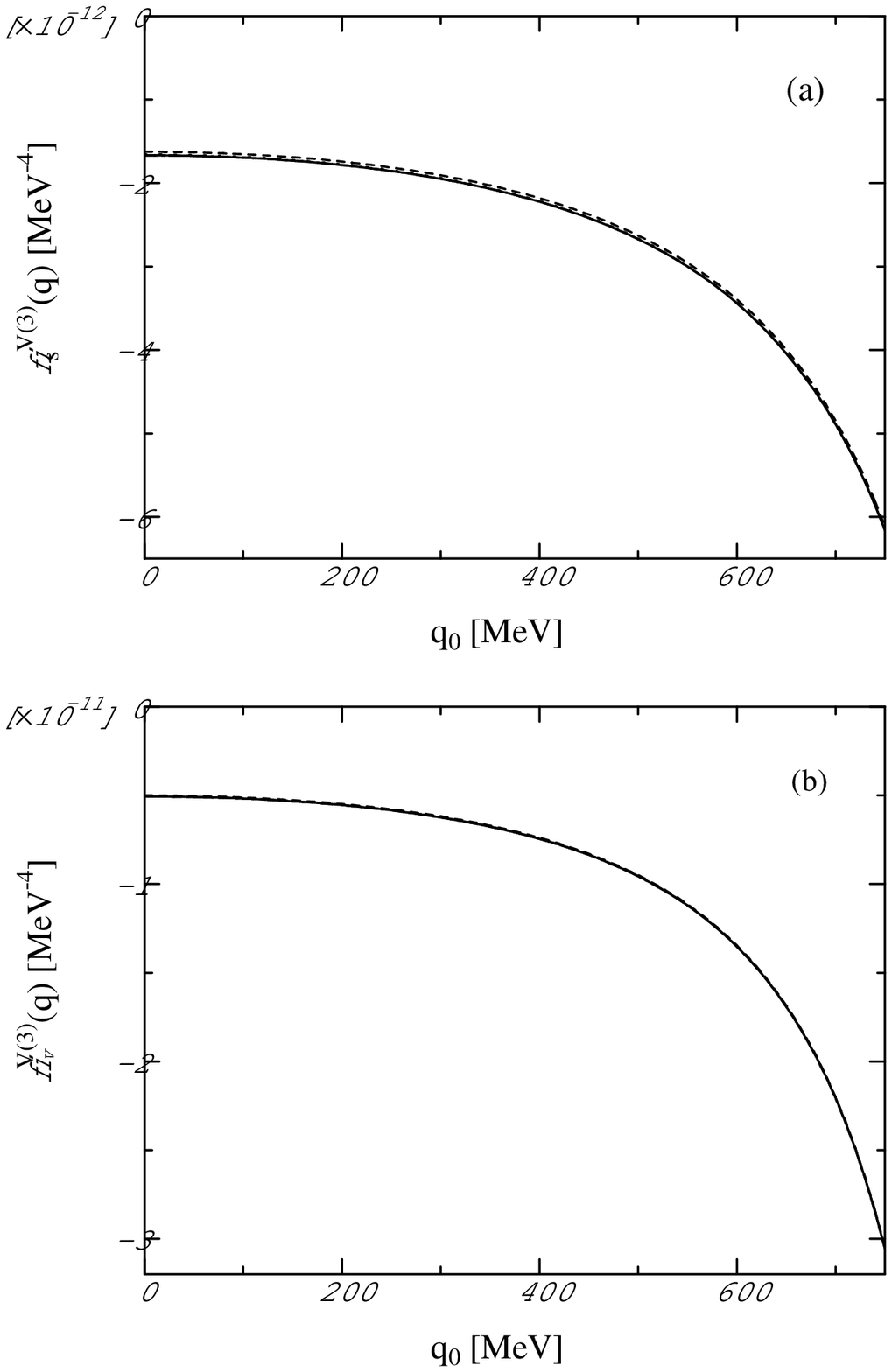}
\caption{
The $\Lambda$-dependence of the third derivatives of $\Pi^V$. 
The meaning of each line is same as in Fig. \ref{2}. 
(a) The $\sigma$-meson self-energy, 
(b) The $\omega$-meson self-energy. 
}
\label{3}
\end{center}
\end{figure}

In Figs. \ref{2} and \ref{3}, we show $\Pi^{V(2)}$ and $\Pi^{V(3)}$ calculated with several cutoff $\Lambda$ = 2, 3, 5 GeV, respectively. 
Naturally, the cutoff $\Lambda$ gives the limit of integration area of
internal momentum in RPA Feynman diagram (Fig. \ref{f}). 
In these results, we see that, although $\Lambda$-dependence of $\Pi^{V(2)}$ is large, the $\Lambda$-dependence of $\Pi^{V(3)}$ is sufficiently small and can be neglected. 

\vspace{1cm}
\begin{center}
  \begin{tabular}{c} 
     \hline Fig.\ref{2} \\ \hline \\
     \hline Fig.\ref{3} \\ \hline
  \end{tabular}
\end{center}
\vspace{1cm}

We also show the results with a finite cutoff $\Lambda_\omega$ which is the upper limit of the integration in the integral Eq.(\ref{eq:dd}). 
The results show the similar features as in Figs. \ref{2} and \ref{3}. 
These results is due to that the integrand in $\Pi^{V(3)}$ decreases 
more rapidly than one in $\Pi^{V(2)}$ as $\omega$-beocomes large and the value of $\Pi^{V(3)}$ is hardly affected by the change of the cutoff. 
In other word, the $\omega$-integration in $\Pi^{V(3)}$ is mainly dominated by the contributions at small $\omega$ which is much smaller than $\Lambda_\omega$. 

\vspace{1cm}
\begin{center}
  \begin{tabular}{c} 
     \hline Fig.\ref{2d} \\ \hline \\
     \hline Fig.\ref{3d} \\ \hline
  \end{tabular}
\end{center}
\vspace{1cm}

Those results indicate that, in QHD with the finite cutoff of order GeV, 
if we want to the physical results which hardly depend on the cutoff,  
we should remove the errors of order $(1/\Lambda^2)$. 

\begin{figure}
\begin{center}
    \includegraphics*[height=16cm]{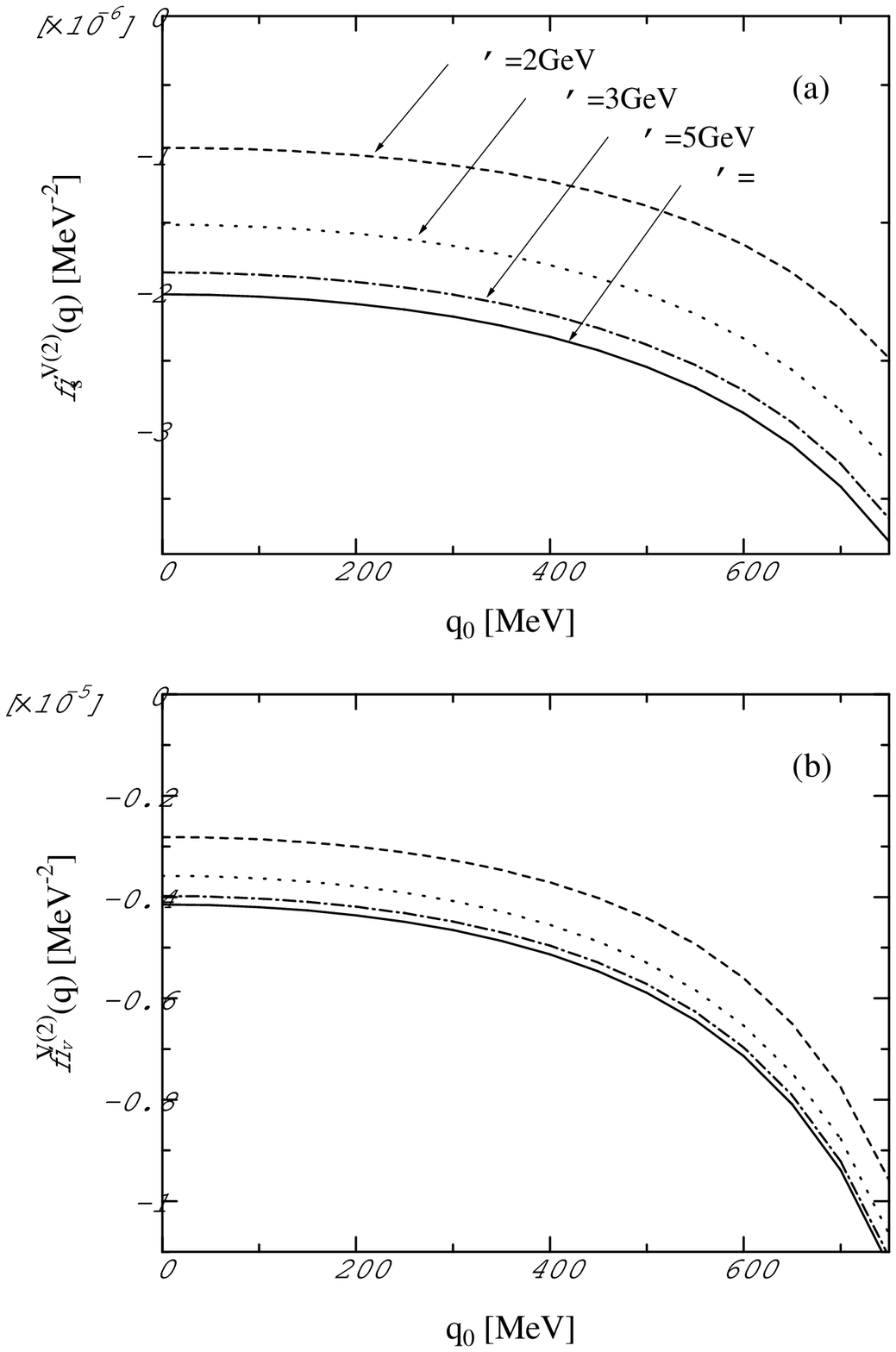}
\caption{
The $\Lambda_{\omega}$ dependence of the second derivative of $\Pi^V$. 
The meaning of each line is same as in Fig. \ref{2}. 
(a) The $\sigma$-meson self-energy, 
(b) The $\omega$-meson self-energy. 
}
\label{2d}
\end{center}
\end{figure}

\begin{figure}
\begin{center}
    \includegraphics*[height=16cm]{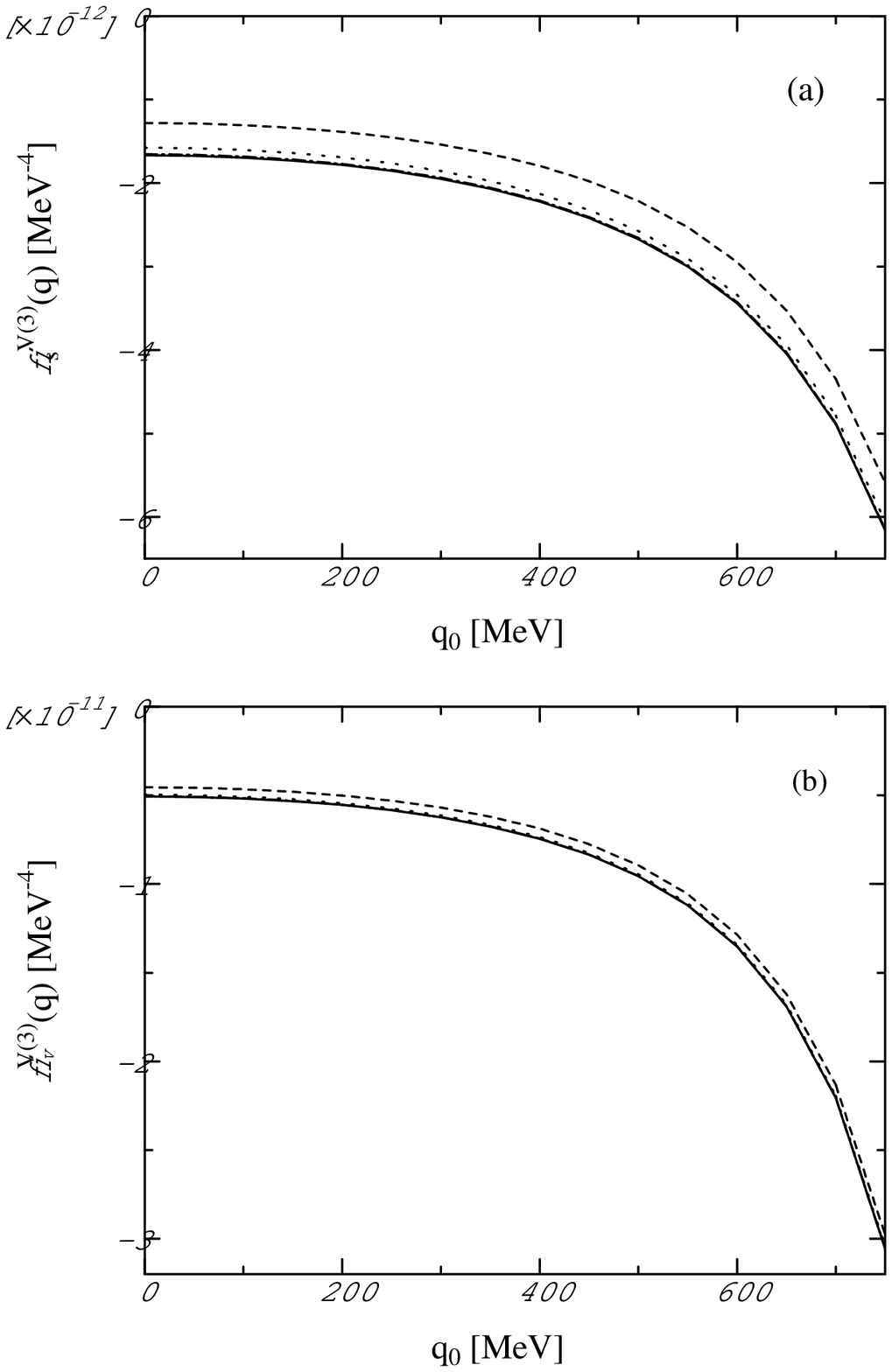}
\caption{
The $\Lambda_{\omega}$ dependence of the third derivative of $\Pi^V$. 
The meaning of each line is same as in Fig. \ref{2}. 
(a) The $\sigma$-meson self-energy, 
(b) The $\omega$-meson self-energy. 
}
\label{3d}
\end{center}
\end{figure}

For example, in the vacuum fluctuation part of RPA $\sigma$-meson
self-energy, we expand around renormalization point $q^2=m_s^2$, $M^*=M$, 
\begin{eqnarray}
\Pi^V_{s}(q,\Lambda) &=& \Pi^V_{s} \bigg|_{{M^*=M \atop q^2=m_s^2}}
+ \frac{\partial \Pi^V_{s}}{\partial M^*} \bigg|_{{M^*=M \atop q^2=m_s^2}}
(M^*-M)
 \nonumber \\
 & & + \frac{\partial \Pi^V_{s}}{\partial q^2} \bigg|_{{M^*=M \atop q^2=m_s^2}}
(q^2-m_s^2)
+ \frac{1}{2}\frac{\partial^2 \Pi^V_{s}}{\partial M^{* 2}}
\bigg|_{{M^*=M \atop q^2=m_s^2}}(M^*-M)^2 \nonumber \\
 & & + \frac{\partial^2 \Pi^V_{s}}{\partial M^* \partial q^2}
\bigg|_{{M^*=M \atop q^2=m_s^2}}(M^*-M)(q^2-m_s^2)
+ \frac{1}{2} \frac{\partial^2 \Pi^V_{s}}{(\partial q^2)^2}
\bigg|_{{M^*=M \atop q^2=m_s^2}}(q^2-m_s^2)^2 \nonumber \\
 & & + \frac{1}{3!}\frac{\partial^3 \Pi^V_{s}}{\partial M^{* 3}}
\bigg|_{{M^*=M \atop q^2=m_s^2}}(M^*-M)^3
+ \frac{1}{2} \frac{\partial^3 \Pi^V_{s}}{\partial M^{* 2} \partial q^2}
\bigg|_{{M^*=M \atop q^2=m_s^2}}(M^*-M)^2(q^2-m_s^2) \nonumber \\
 & & + \frac{1}{4!}\frac{\partial^4 \Pi^V_{s}}{\partial M^{* 4}}
\bigg|_{{M^*=M \atop q^2=m_s^2}}(M^*-M)^4
+\cdot \cdot \cdot \cdot \cdot \cdot \cdot.
\label{eq:pe}
\end{eqnarray}
Here, in the limit $\Lambda \rightarrow \infty$, 
the first four terms of the right hand side (r. h. s.) in Eq. (\ref{eq:pe}) are diverging. 
In the ordinary renormalization procedure, 
we must determine only these divergent terms phenomenologically, and
we can $calculate$ the remaining terms. 
However, since the cutoff is finite in QHD, the cutoff dependence of the next five terms in r. h. s. of Eq. (\ref{eq:pe}) is not small. 
The figures \ref{2d} and \ref{3d} indicates that 
we can not ignore the error of O(1/$\Lambda^2$) 
in those five terms. 
Therefore, we should determine 9 terms which has the errors greater than O(1/$\Lambda^2$) 
phenomenologically. 

\begin{figure}
\begin{center}
    \includegraphics*[height=16cm]{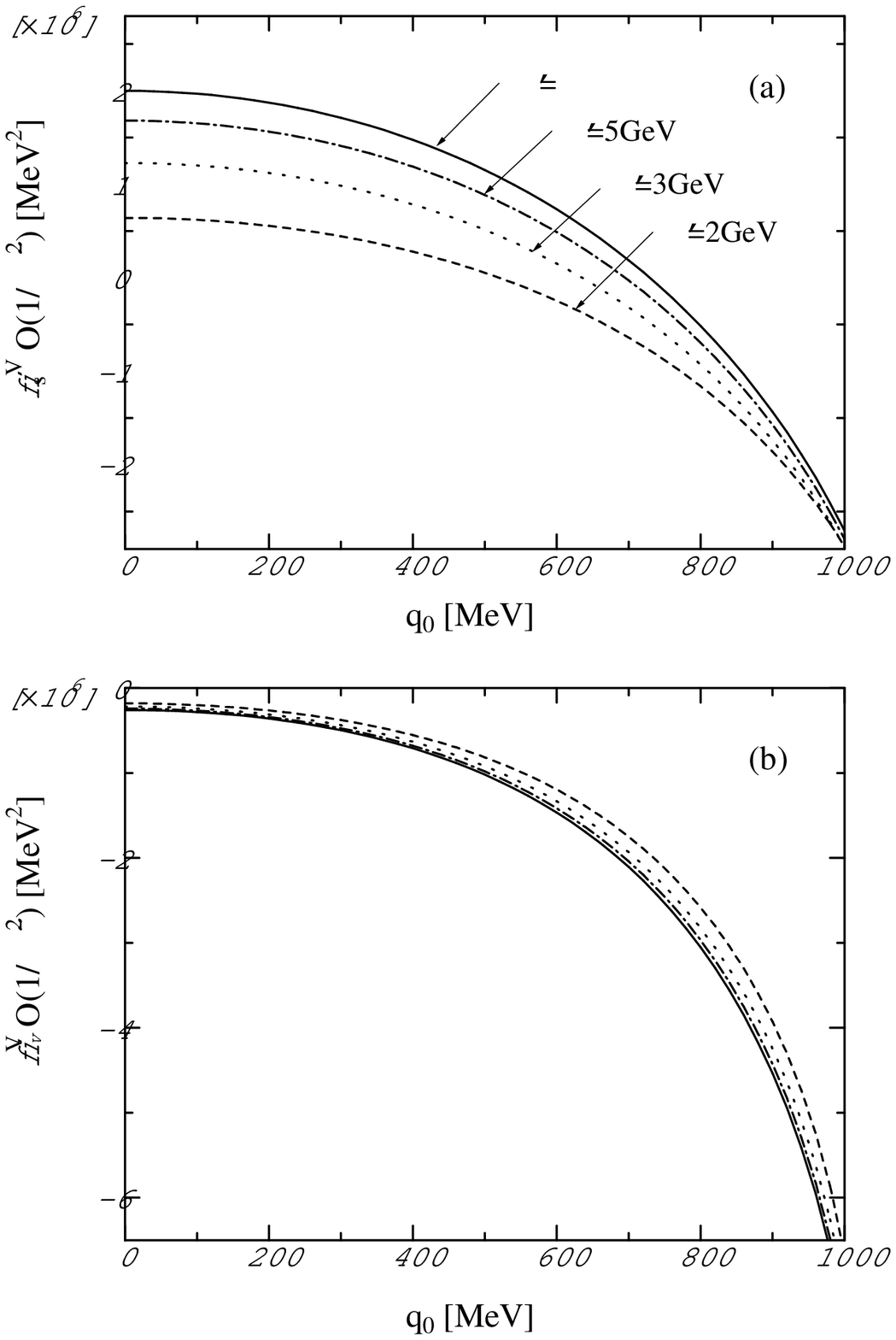}
\caption{
The $\Lambda$-dependence of $\Pi^V_{O(1/\Lambda^2)}$. 
The meaning of each line is same as in Fig. \ref{2}. 
(a) The $\sigma$-meson self-energy, 
(b) The $\omega$-meson self-energy. 
}
\label{ln}
\end{center}
\end{figure}

\begin{figure}
\begin{center}
    \includegraphics*[height=16cm]{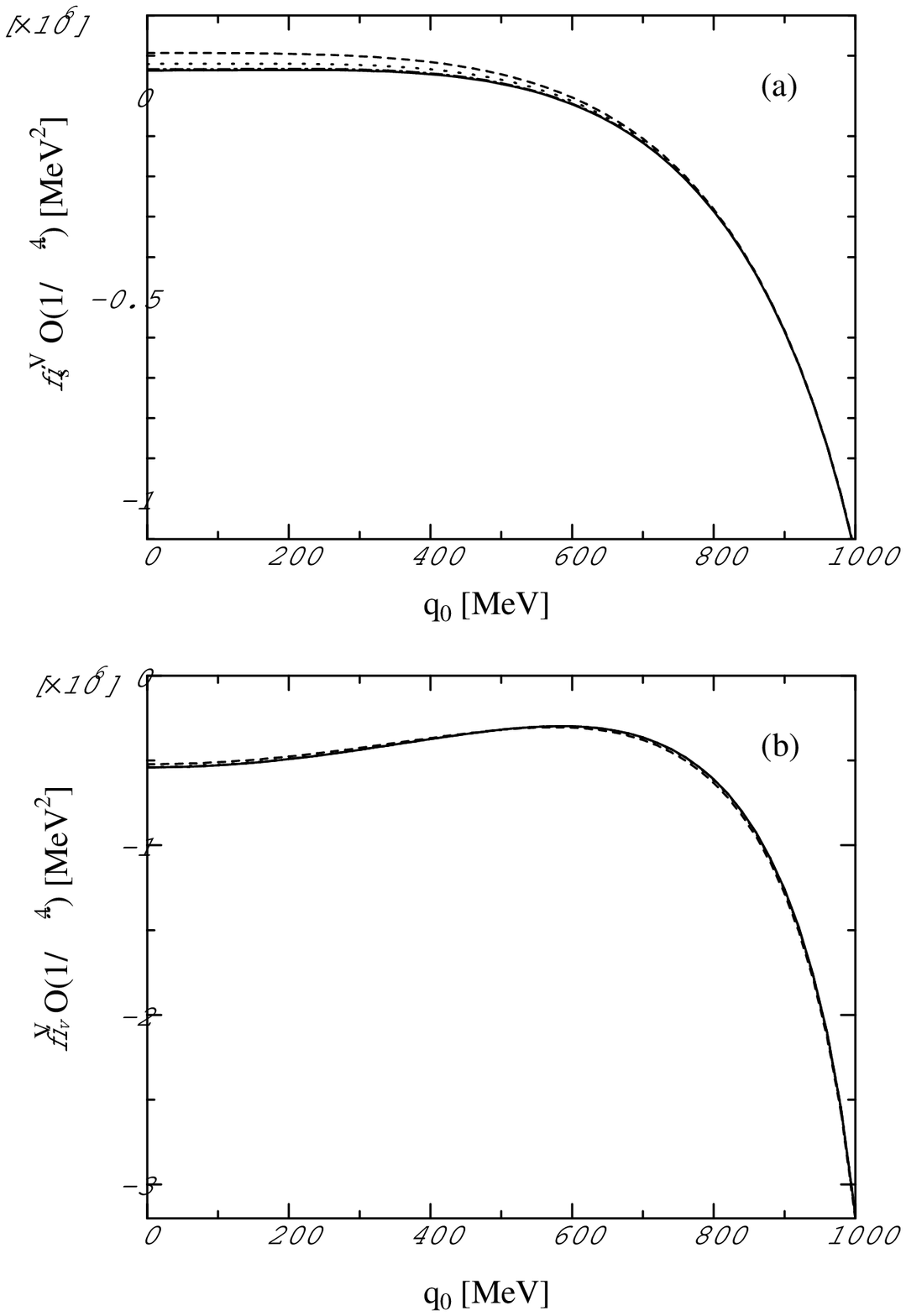}
\caption{
The $\Lambda$-dependence of $\Pi^V_{O(1/\Lambda^4)}$. 
The meaning of each line is same as in Fig. \ref{2}. 
(a) The $\sigma$-meson self-energy, 
(b) The $\omega$-meson self-energy. 
}
\label{l2}
\end{center}
\end{figure}

Finally, in Fig. \ref{ln}, we show the cutoff dependence of $\Pi^V_{O(1/\Lambda^2)}$ from which the first four terms in Eq. (\ref{eq:pe}) are subtracted. 
In Fig. \ref{l2}, we also show the cutoff dependence of $\Pi^V_{O(1/\Lambda^4)}$ from which the first nine terms in Eq.(\ref{eq:pe}) are substracted. 
As expected, we see again that we can not ignore the errors of O(1/$\Lambda^2$). 
In QHD, we must determine much more terms phenomenologically than in the usual renormalization procedure. 

\vspace{1cm}
\begin{center}
  \begin{tabular}{c} 
     \hline Fig.\ref{ln} \\ \hline \\
     \hline Fig.\ref{l2} \\ \hline
  \end{tabular}
\end{center}
\vspace{1cm}

\section{Summary}

The results obtained in this paper are summarized as follows. 

(1) We formulated the method of renormalization in the vacuum polarization effects in QHD as the finite cutoff field theory.
We can remove the errors which arise from the finiteness and ambiguity of the cutoff, order by order, by determining the coefficients of the effective Lagrangian phenomenologically.

(2) We reformulated the method by using a differential equation in the finite cutoff field theory. 
In this reformulation, it is more clear that the physical result hardly depends on 
the cutoff $\Lambda$ and the details of its regulator which are introduced ad hoc. 
In this method, the $\Lambda$-dependence of physical quantity becomes small, by differentiating the physical quantity which has large $\Lambda$-dependence with respect to the external parameters. 
After that, we integrate the obtained differential equation by the external parameter. 
The cutoff dependence is removed by the phenomenological determinations of the integration constants which appear in the integration. 

(3) We ascertained dispersion relations of RPA $\sigma$ and $\omega$ self-energies at finite density. 
In vacuum fluctuation part which diverges in the limit $\Lambda \rightarrow \infty$, we examined a dispersion relations in the conversing derivatives of $\Pi^V$ rather than $\Pi^V$ itself. 
It is shown that such a derivatives well preserves the dispersion relations, although $\Pi^V$ itself is diverging. 

(4) We examined the dependence on the finite cutoff in the meson self-energies numerically, 
and discussed an utility of our renormalization method in QHD.  
We found that the errors of O(1/$\Lambda^2$) which arise from the finiteness of
cutoff should not be ignored in QHD where $\Lambda$ is order GeV. 
This result indicates that we must determine much more terms in the effective QHD Lagrangian phenomenologically than in the usual renormalization procedure. 

It is very interesting to calculate the effective meson masses at finite baryon density by using our method of renormalization. 
It is in progress.


\begin{center}
{\large  \bf Acknowledgements}

\end{center}

The authors would like to thank T. Kohmura, T. Suzuki, M. Yahiro, K. Harada,
K.-I. Aoki, H. Yoneyama and T. Mitsumori for useful discussions and
suggestions.
This work is supported in part by the Japanese Scientific Foundation
(Grant No. 09740208).


\begin{center}
{\large  \bf Appendix}
\end{center}

Here we derive the dispersion relation in the case of the particle-hole part of the $\sigma$-meson self-energy. 
We can derive the relation for the other parts in similar way. 

The $\sigma$-meson self-energy in the RPA calculations is given by 
\begin{equation}
\Pi_s(q) = -ig^{2}_{s}\int\frac{d^{4}k}{(2\pi)^4}
Tr[G(k)G(k+q)]
\label{eq:App1}
\end{equation}
For this purpose, it is convenient to use the particle-hole-antiparticle decomposition \cite{rf:Liu}-\cite{rf:Nakano2} of the nucleon propagator rather than the Feynman-density decomposition. 
\begin{eqnarray}
G(k) &=& (\gamma^{\mu}k^{\ast}_{\mu}+M^{\ast}) \frac{1}{2E^{\ast}(k)}
\nonumber\\
 & & \times \Bigl[\frac{1-\theta(k_{F}-|\mbox{\boldmath $k$}|)}
{k_0^*-E^{\ast}(k)+i\epsilon}
+\frac{\theta(k_{F}-|\mbox{\boldmath $k$}|)}
{k_0^*-E^{\ast}(k)-i\epsilon}
-\frac{1}{k_0^*+E^{\ast}(k)-i\epsilon} \Bigr] \nonumber \\
&\equiv& G_{p}(k)+G_{h}(k)+G_{a}(k), 
\label{eq:App2}
\end{eqnarray}
where $G_p$, $G_h$ and $G_a$ represent the particle, the hole and the antiparticle propagations, respectively. 
The particle-hole part of the $\sigma$-meson self-energy is given by 
\begin{eqnarray}
\Pi^{ph}_s(q) &=& -ig^{2}_{s}\int\frac{d^{4}k}{(2\pi)^4}
Tr[G_p(k)G_h(k+q)+G_h(k)G_p(k+q)] \nonumber \\
&\equiv & \Pi_+(q)+\Pi_-(q), 
\label{eq:App3}
\end{eqnarray}
where 
\begin{eqnarray}
\Pi_+(q)
&=& -ig^{2}_{s}\int\frac{d^{4}k}{(2\pi)^4}
\frac{M^{*2}+k \cdot (k+q)}{E^*(k)E^*(k+q)}
\nonumber \\
& & \times \biggl \{ \theta(k_{F}-|\mbox{\boldmath $k$}|)
[1- \theta(k_{F}-|\mbox{\boldmath $k$}+\mbox{\boldmath $q$}|)]
\frac{1}{[k_0-E^*(k)-i \epsilon]}\frac{1}{[k_0+q_0-E^*(k+q)+i \epsilon]} 
\biggr \}
\nonumber\\
&=& \frac{g^{2}_{s}}{2}\int\frac{d^{3}k}{(2\pi)^3}
\frac{\theta(k_{F}-|\mbox{\boldmath $k$}|)
[1- \theta(k_{F}-|\mbox{\boldmath $k$}+\mbox{\boldmath $q$}|)]}
{E^*(k)E^*(k+q)} \nonumber \\
& & \times \biggl \{ 
2(E^*(k)+E^*(k+q))+(4M^{*2}-q^2)
\nonumber \\
& & \times \Bigl[\frac{1}{E^*(k)+q_0-E^*(k+q)+i \epsilon} \Bigr] \biggr \}. 
\label{eq:App4} 
\end{eqnarray}
and 
\begin{eqnarray}
\Pi_-(q) 
&=& -ig^{2}_{s}\int\frac{d^{4}k}{(2\pi)^4}
\frac{M^{*2}+k \cdot (k+q)}{E^*(k)E^*(k+q)}
\nonumber \\
& & \times \biggl\{ 
\theta(k_{F}-|\mbox{\boldmath $k$}+\mbox{\boldmath $q$}|)
[1- \theta(k_{F}-|\mbox{\boldmath $k$}|)]
\frac{1}{[k_0+q_0-E^*(k+q)-i \epsilon]}\frac{1}{[k_0-E^*(k)+i \epsilon]}
\biggr \}   
\nonumber\\
&=& \frac{g^{2}_{s}}{2}\int\frac{d^{3}k}{(2\pi)^3}
\frac{\theta(k_{F}-|\mbox{\boldmath $k$}|)
[1- \theta(k_{F}-|\mbox{\boldmath $k$}+\mbox{\boldmath $q$}|)]}
{E^*(k)E^*(k+q)} \nonumber \\
& & \times \biggl \{ 2(E^*(k)+E^*(k+q))+(4M^{*2}-q^2)
\nonumber \\
& & \times \Bigl[ \frac{1}{E^*(k)-q_0-E^*(k+q)+i \epsilon}
\Bigr] \biggr \}. 
\label{eq:App5} 
\end{eqnarray}
Using the relation, 
\begin{eqnarray}
\frac{1}{x \pm i \epsilon}={\bf P }\frac{1}{x} \mp i \pi \delta(x)
\label{eq:App6}
\end{eqnarray}
we get 
\begin{eqnarray}
Im \Pi_+(q) &=& -g^{2}_{s} \frac{\pi}{2}(4M^{*2}-q^2)
\int\frac{d^{3}k}{(2\pi)^3}
\frac{\theta(k_{F}-|\mbox{\boldmath $k$}|)
[1- \theta(k_{F}-|\mbox{\boldmath $k$}+\mbox{\boldmath $q$}|)]}
{E^*(k)E^*(k+q)} \nonumber \\
& & \times \delta(E^*(k)+q_0-E^*(k+q))
\label{eq:App7} 
\end{eqnarray}
and 
\begin{eqnarray}
Im \Pi_-(q) &=& -g^{2}_{s} \frac{\pi}{2}(4M^{*2}-q^2)
\int\frac{d^{3}k}{(2\pi)^3}
\frac{\theta(k_{F}-|\mbox{\boldmath $k$}|)
[1- \theta(k_{F}-|\mbox{\boldmath $k$}+\mbox{\boldmath $q$}|)]}
{E^*(k)E^*(k+q)} \nonumber \\
& & \times \delta(E^*(k)-q_0-E^*(k+q)). 
\label{eq:App8}
\end{eqnarray}
From these equations, we see that $Im \Pi_-(q_0,{\bf q})=Im \Pi_+(-q_0,{\bf q})$. 

If $\vert {\bf k}+{\bf q}\vert \leq \vert {\bf k}\vert$, 
there is no contribution for the $k$-integral in Eq. (\ref{eq:App7}). 
On the other hand, if $\vert {\bf k}+{\bf q}\vert > \vert {\bf k}\vert$, $E^*(k)-E^*(k+q)<0$. 
Therefore, we get 
\begin{equation} 
Im \Pi_+(q)=0~~~~~{\rm for}~~~~~q_0\leq 0. 
\label{eq:App9a}
\end{equation}
Similarly, we get 
\begin{equation} 
Im \Pi_-(q)=0~~~~~{\rm for}~~~~~q_0\geq 0. 
\label{eq:App9b}
\end{equation}

Since $\Pi_+$ has no pole in the upper half-plane of the complex $q_0$-plane, we get the dispersion relation 
\begin{equation}
Re\Pi_+(q)=\frac{{\rm \bf P}}{\pi}\int_{-\infty}^\infty
\frac{Im\Pi_+(\omega,\mbox{\boldmath $q$})}
{\omega-q_0}d\omega. 
\label{eq:App10}
\end{equation}
Similarly, since $\Pi_-$ has no pole in the lower half-plane of the complex $q_0$-plane, we get the dispersion relation 
\begin{equation}
Re\Pi_-(q)=-\frac{{\rm \bf P}}{\pi}\int_{-\infty}^\infty
\frac{Im\Pi_-(\omega,\mbox{\boldmath $q$})}
{\omega-q_0}d\omega. 
\label{eq:App11}
\end{equation}
Therefore, 
we get 
\begin{eqnarray}
Re\Pi_s^{ph}&=&Re\Pi_+(q)+Re\Pi_-(q)
\nonumber\\
&=&\frac{{\rm \bf P}}{\pi}\int_{-\infty}^\infty
\frac{Im\Pi_+(\omega,\mbox{\boldmath $q$})}
{\omega-q_0}d\omega 
-\frac{{\rm \bf P}}{\pi}\int_{-\infty}^\infty
\frac{Im\Pi_-(\omega,\mbox{\boldmath $q$})}
{\omega-q_0}d\omega
\nonumber\\
&=&\frac{{\rm \bf P}}{\pi}\int_0^{\infty}
\frac{Im\Pi_+(\omega,\mbox{\boldmath $q$})}
{\omega-q_0}d\omega 
-\frac{{\rm \bf P}}{\pi}\int_{-\infty}^0
\frac{Im\Pi_+(-\omega,\mbox{\boldmath $q$})}
{\omega-q_0}d\omega
\nonumber\\
&=&\frac{{\rm \bf P}}{\pi}\int_0^{\infty}
\frac{Im\Pi_+(\omega,\mbox{\boldmath $q$})}
{\omega-q_0}d\omega 
-\frac{{\rm \bf P}}{\pi}\int_0^{\infty}
\frac{Im\Pi_+(\omega,\mbox{\boldmath $q$})}
{-\omega-q_0}d\omega
\nonumber\\
&=&\frac{2{\rm \bf P}}{\pi}\int_0^\infty
\frac{Im\Pi^{ph}_s(\omega,\mbox{\boldmath $q$})\cdot\omega}
{\omega^2-q_0^2}d\omega,
\label{eq:App12}
\end{eqnarray}
The dispersion relations for the other parts of $\Pi$ can be also shown in similar way. 
For example, if we use the Feyman propagator $G_F(k)$ in $\Pi$, 
we get the dispersion relation for $\Pi^V$. 
In that case, we easily see that 
the condition for the dispersion relation in $q_0$-plane can be consistent with the condition for the Wick rotation in $k_0$-plane, since we decompose $\Pi^V$ into $\Pi^V_+$ and $\Pi^V_-$ to apply the dispersion relation. 

\vspace{2cm}



\begin{figure}
\begin{center}
    \includegraphics*[height=9cm]{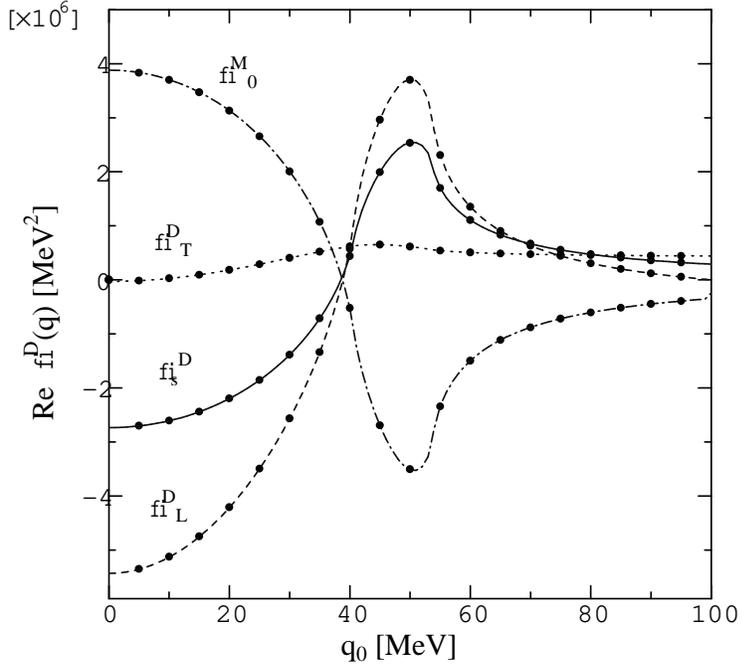}
\caption{
Dispersion relation for the density part. 
The solid, dotted, dashed and dashed-dotted lines are 
$Re\Pi^D_s$, $Re\Pi^D_T$,$Re\Pi^D_L$ and $Re\Pi^M_0$, 
where $\Pi^D_T \equiv \Pi^D_{00}-\Pi^D_{11}$
and $\Pi^D_L \equiv \Pi^D_{22}=\Pi^D_{33}$ as $\mbox{\boldmath $q$}=(q,0,0)$. 
The corresponding results calculated by the dispersion relation Eq. (\ref{eq:dd}) are shown by filled circles. 
We put $q$ = 100 MeV in these calculation.  
}
\label{dd}
\end{center}
\end{figure}

\end{document}